\documentclass[a4paper,11pt]{article}
\pdfoutput=1
\usepackage{jheppub}
\usepackage{amsfonts,amssymb,amsmath}
\usepackage{color}
\usepackage{graphicx}
\usepackage{subcaption}
\usepackage[normalem]{ulem}

\graphicspath{{./Images/}}

\newcommand{\tcr}{\textcolor{red}}
\newcommand{\tcb}{\textcolor{blue}}

\newcommand{\be}{\begin{equation}}
\newcommand{\ee}{\end{equation}}
\newcommand{\ba}{\begin{eqnarray}}
\newcommand{\ea}{\end{eqnarray}}
\newcommand{\nn}{\nonumber}
\newcommand{\lb}{\left(}
\newcommand{\rb}{\right)}
\newcommand{\qbra}[1]{\left< #1 \right|}
\newcommand{\qket}[1]{\left| #1 \right>}

\title{Breakdown of the Equal Area Law for Holographic Entanglement Entropy}

\author[a,b]{Fiona McCarthy}
\author[a,b]{David Kubiz\v n\'ak}
\author[b,a]{Robert B. Mann}

\affiliation[a]{Perimeter Institute for Theoretical Physics, 31 Caroline Street North, Waterloo,
ON, N2L 2Y5, Canada}
\affiliation[b]{Department of Physics and Astronomy, University of Waterloo,
Waterloo, Ontario, Canada, N2L 3G1}

\emailAdd{fmccarthy@perimeterinstitute.ca}
\emailAdd{dkubiznak@perimeterinstitute.ca}
\emailAdd{rbmann@uwaterloo.ca}

\abstract{We  investigate a holographic version of Maxwell's equal area law analogous to that for the phase transition in the black hole temperature/black hole entropy plane of a charged AdS black hole.  We consider proposed area laws for both  the black hole temperature/holographic entanglement entropy plane and the black hole temperature/2-point correlation function plane.  Despite recent claims to the contrary, we demonstrate numerically that neither proposal is valid.
 We argue that there is no physical reason to expect such a construction in these planes.
}

\preprint{}

\begin{document}

\maketitle


\section{Introduction}

Maxwell's equal area law, which states that two phases coexist when the areas above and below a line of constant pressure $P$ drawn through a pressure/volume curve are equal, is one of the hallmarks of thermodynamics.  It provides a straightforward computational method for obtaining the coexistence boundary between any two phases (separated by a first order phase transition), and generalizes straightforwardly to any pair of conjugate thermodynamic variables.  In recent years it has found utility in the thermodynamics of AdS black holes \cite{Spallucci:2013osa,Lan:2015bia, Xu:2015hba}, where the magnitude of the cosmological constant is interpreted as thermodynamic pressure and the conjugate volume $V$ is obtained by
differentiating the black hole mass with respect to pressure \cite{Kubiznak:2016qmn}.

Recently there has been interest in defining equal area laws for holographic entanglement entropy \cite{Nguyen:2015wfa, Zeng:2015tfj, Sun:2016til, Liu:2017jbm}
as well as two-point correlation functions
\cite{Zeng:2015wtt, Zeng:2016sei, Mo:2016cmi,Mo:2016ijb, Zeng:2016aly, Zeng:2016fsb, ElMoumni:2016eqh, Li:2017gyc, Li:2017xiv, Zeng:2017zlm}.
These equal area laws have been studied for spacetimes dual to AdS black holes with phase transitions obeying an equal area law in the black hole temperature ($T$)/black hole entropy ($S$) plane such as the charged AdS black hole undergoing a first-order phase transition \cite{Chamblin:1999tk, Chamblin:1999hg,Kubiznak:2012wp}. The holographic equal area laws have been considered in both  the $T$/entanglement entropy plane and the $T$/geodesic length plane, as the two-point correlation function is given by the exponential of geodesic length \cite{Balasubramanian:1999zv}. It has been claimed in a number of cases \cite{Nguyen:2015wfa, Zeng:2015tfj,Liu:2017jbm,Zeng:2015wtt, Zeng:2016sei, Mo:2016cmi,Mo:2016ijb, Zeng:2016aly, Zeng:2016fsb, ElMoumni:2016eqh, Li:2017gyc, Li:2017xiv, Zeng:2017zlm}
that isocharges in these planes obey Maxwell's equal area construction at the phase transition temperature of the black hole $T_*$, such that the areas bounded above and below the isocharge and the isotherm $T=T_*$ are equal, just as is true for black hole temperature and entropy.

The similarity between holographic entanglement entropy (HEE) and black hole entropy \cite{Johnson:2013dka} motivated the  idea of an equal area law for holographic entanglement entropy, where it was first claimed  \cite{Nguyen:2015wfa} that there is numerical evidence for an HEE equal area law for the
(near critically) charged AdS black hole in $3+1$ dimensions. It was further claimed that this equal area law for HEE sharpened the similarity between black hole entropy and HEE. However, numerical evidence that the equal area law for HEE breaks down was subsequently presented \cite{Sun:2016til}, the discrepancy growing as isocharges are chosen further away from their originally considered  \cite{Nguyen:2015wfa} near-critical values.   More recently, claims that a holographic equal area law holds in the $T$/geodesic length plane have appeared  \cite{Zeng:2015wtt, Zeng:2016sei, Mo:2016cmi,Mo:2016ijb, Zeng:2016aly, Zeng:2016fsb, ElMoumni:2016eqh, Li:2017gyc, Li:2017xiv, Zeng:2017zlm}.

Here we present the results of an investigation into both proposals for a holographic equal area law.  We find that any claim of an equal area law holding in either the $T$/entanglement entropy or $T$/geodesic length plane is untrue and unfounded.  We find numerically that such equal area laws are not satisfied in either case, and  explain how such erroneous claims could arise.
Furthermore, we point out that there is no reason to expect this based on an appropriate consideration of the relevant thermodynamics.


\section{Phase structure of charged AdS black holes}\label{sec:phase_sctructure}

It is well known \cite{Chamblin:1999tk, Chamblin:1999hg,Kubiznak:2012wp} that in a canonical (fixed charge) ensemble the thermodynamics
of charged AdS black holes features a first order (small black hole/large black hole) phase transition, with the corresponding thermodynamics governed by
the black hole free energy.
Alternatively, one can describe such a phenomenon using the Maxwell equal area construction in the $T-S$ (and/or $P-V$) planes. Since the charged AdS black hole will serve as a testground for our investigation of validity of the holographic equal area laws, let us start by briefly recapitulating these bulk results.

 A $d$-dimensional charged AdS black hole is a solution to the
Einstein--Maxwell anti de Sitter action  \cite{Chamblin:1999tk}
\be
I=-\frac{1}{16\pi G}\int d^dx \sqrt{-g} \left[ R-F^2+\frac{(d-1)(d-2))}{l^2}\right]\,,
\ee
where $l$ is the AdS length scale, given by the following metric:
\begin{gather}\label{RN_AdS_Metric}
ds^2=-f(r)dt^2+\frac{dr^2}{f(r)}+r^2d\Omega_{(d-2)}^2,\\
f(r)=1-\frac{m}{r^{d-3}}+\frac{q^2}{r^{2(d-3)}}+\frac{r^2}{l^2}\,.\label{f_r}
\end{gather}
The parameters $m$ and $q$ are related to the ADM mass and charge of the black hole $M$ and $Q$ via
\begin{align}
M&=\frac{(d-2)\, \omega_{(d-2)}}{16\pi G}m\,,\label{ADM_M}\\
Q&=\frac{\sqrt{2(d-2)(d-3)}\,\omega_{(d-2)}}{8 \pi G}q\,, \label{ADM_Q}
\end{align}
where $\omega_{(d-2)}$ is the area of the unit $(d-2)$-sphere $\omega_{(d-2)}=\frac{2\pi^{\frac{d-1}{2}}}{\Gamma\lb\frac{d-1}{2}\rb}$.
The temperature $T=\frac{f'(r_+)}{4\pi}$ and entropy $S = \omega_{(d-2)} r^{d-2}_+ /4 $ are straightforwardly computed.
The solution for the gauge potential is
\be
A=\lb-\frac{1}{c}\frac{q}{r^{d-3}}+\Phi\rb dt\,,
\ee
where $c=\sqrt{\frac{2(d-3)}{d-2}}$, and $\Phi$ is a constant.  Choosing $\Phi=\frac{1}{c}\frac{q}{r_+^{d-3}}$, with $r_+$ the horizon radius of the black hole, the potential $A$ vanishes on the horizon. The above black hole quantities obey the following standard first law of black hole thermodynamics:
\be\label{dM}
dM=T dS+\Phi dQ\,.
\ee

Specializing to $d= 3+1$ dimensions,  we can express $T$  as a function of $S$ and $Q$
\be\label{TSQ}
T(S,Q)=\frac{1}{4\pi}\lb \frac{3}{l^2}\sqrt{\frac{S}{\pi}}+\sqrt{\frac{\pi}{S}}-Q^2\frac{\pi^{\frac{3}{2}}}{S^{\frac{3}{2}}}\rb
\ee
and also obtain in the canonical ensemble
\be
F=M-TS=\frac{1}{4 l^2}\lb  l^2 r_+-r_+^3+\frac{3 Q^2l^2}{r_+}\rb	
\ee
for the free energy $F=M-TS$, which completely governs the thermodynamic behavior of the bulk black hole.

\begin{figure}[h!]
\begin{center}
\includegraphics[scale=0.75]{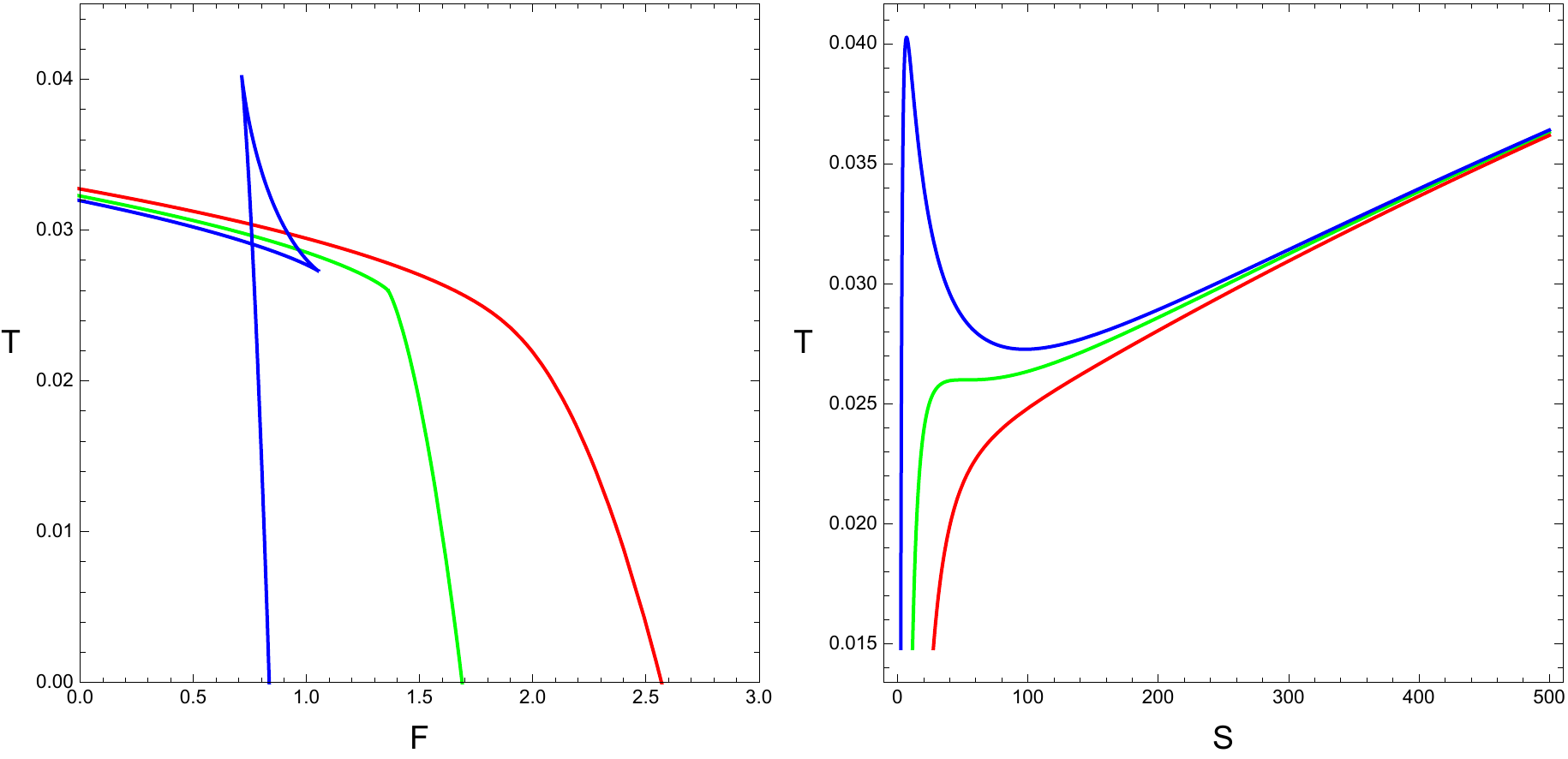}
\end{center}
\caption{\textbf{  Phase transition of a bulk black hole.}
On the left, the behavior of the temperature $T$ against the free energy $F$ of a charged black hole in $d=4$ is shown. On the right is the behavior of $T$ against the entropy $S$. In each case we have plotted the critical isocharges (green), and isocharges at $Q>Q_{crit}$ (red) and $Q<Q_{crit}$ (blue). Below criticality we see the swallowtail behavior of $F$,  characteristic of a first order phase transition, and the oscillatory behaviour of $T$; at criticality the swallowtail becomes a cusp in the $T-F$ plane, and an inflection point in the $T-S$ plane--- the phase transition is here of second order.  The AdS radius $l$ has been set to 10 for which the critical charge is $Q_c=10/6$. {The values of $Q$ on these isocharges are $Q=0.5\, Q_{crit}$ (blue), $Q=Q_{crit}$ (green) and $Q=1.5\, Q_{crit}$ (red).}
}
\label{fig:phase_behavior}
\end{figure}

Namely, when isocharge lines are plotted in the $F-T$ plane, swallowtail behaviour characteristic of a first-order phase transition is observed for sufficiently small charges \cite{Chamblin:1999tk,Chamblin:1999hg,Kubiznak:2012wp} (Fig.~\ref{fig:phase_behavior}).
The phase transition temperature $T_*$ occurs at a point at which the derivatives of the global minimum of $F$ become discontinuous, that is, at a point
where the swallowtail intersects itself. As charge increases, the swallowtail diminishes and eventually terminates at a critical point characterized by $Q=Q_{crit}$ and $T=T_{crit}$ at which the phase transition becomes second order. For $Q>Q_{crit}$ the swallowtail no longer exists and only one phase of black holes is present.

\begin{figure}[h!]
\begin{center}
\includegraphics[scale=0.6]{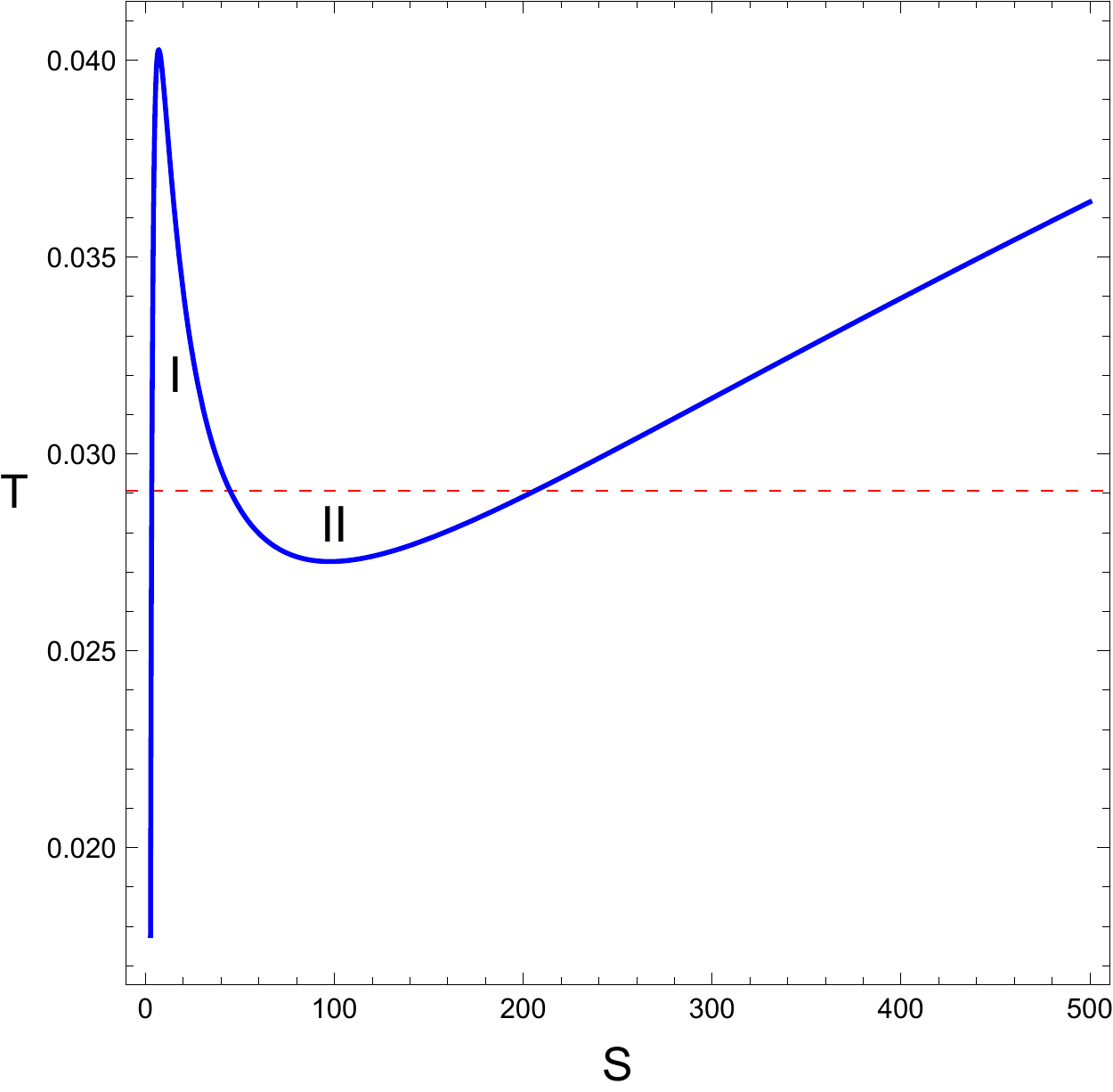}
\caption{\textbf{ Maxwell's equal area law in the bulk.} The phase transition temperature $T=T_*$ is the one at which Areas I and II bounded by the isocharge curve in the $(T,S)$ plane are equal.}
\label{fig:areas}
\end{center}
\end{figure}

The thermodynamic behaviour can alternatively be inferred by studying  isocharge lines in the $T-S$ plane.
Namely, when $T$ is plotted against $S$ for corresponding values of $Q$ (right Fig.~\ref{fig:phase_behavior}), we see that the swallowtail corresponds to an oscillatory behavior in $T$, and the disappearance of the swallowtail at $Q=Q_{crit}$ corresponds to a point of inflection in $T$.  In particular, the critical point quantities $Q_{crit}, S_{crit}, T_{crit}$ can be found by solving explicitly for the inflection point
\be
\frac{\partial T}{\partial S}=\frac{\partial ^2T}{\partial S^2}=0\,,
\ee
together with \eqref{TSQ}, while the phase transition temperature $T_*$ (for $Q<Q_{crit})$ is determined from
 Maxwell's \textit{equal area} construction \cite{Chamblin:1999tk}:
\be\label{integration_areas}
\int_{S_1}^{S_2} T(S,Q) dS-T_*(S_2-S_1)=T_* (S_3-S_2)-\int _{S_2}^{S_3}
T(S,Q) dS\,,
\ee
with $S_1$, $S_2$, $S_3$ given by the solutions of $T(S,Q)=T_*$ in ascending order. Graphically, this corresponds to
\be\label{equal_area_law}
\mathrm{Area}(\mathrm{I})=\mathrm{Area}(\mathrm{II}),
\ee
with Area(I) and Area(II) the areas bounded above and below by $T(S,Q)$ and $T_*$, as depicted in Fig. \ref{fig:areas}.

It is easy to see that   Maxwell's equal area law directly follows from the first law for the free energy:
\begin{align}
dF=-S dT+\Phi dQ\,,
\end{align}
which is a Legendre equivalent of \eqref{dM}. As $dF$ is an exact differential, we have the equal area condition:
\be
\oint SdT=0\quad \Rightarrow \quad
 T_*\lb S_3-S_1\rb =\int_{S_1}^{S_3}T dS\,,\label{maxwell_larger_areas}
\ee
on an isocharge ($dQ=0$),
with $T_*$ the temperature of the phase transition isotherm, c.f. Eq.~\eqref{integration_areas}.  Of course, the same derivation of the equal area law applies when the $T-S$ plane is (for example) replaced by the $P-V$ plane of the extended phase space thermodynamics \cite{Kubiznak:2016qmn}.

\section{Testing holographic equal area laws}

The qualitatively similar behavior of HEE and black hole entropy when plotted against black hole temperature \cite{Johnson:2013dka}, cf. Fig.~\ref{fig:phase_behavior} and Fig.~\ref{fig:entanglement_behavior},
have motivated investigations of potentially interesting phase structure in the QFT dual to a charged AdS black hole.  Both
entanglement entropy \cite{Nguyen:2015wfa, Zeng:2015tfj, Sun:2016til,Zeng:2016aly, Li:2017xiv, Liu:2017jbm,Zeng:2015wtt, Zeng:2016sei,Mo:2016ijb, Zeng:2016fsb, ElMoumni:2016eqh, Li:2017gyc, Zeng:2017zlm} and  two-point correlation functions \cite{Zeng:2015wtt, Zeng:2016sei, Mo:2016cmi,Mo:2016ijb, Zeng:2016aly, Zeng:2016fsb, ElMoumni:2016eqh, Li:2017gyc, Li:2017xiv, Zeng:2017zlm} have been considered to this end.  In both cases  equal area constructions have been respectively proposed in the \textit{black hole temperature}/\textit{entanglement entropy} plane and the \textit{black hole temperature}/\textit{two point corelation function} plane, where a constant entangling region or pair of points are chosen on the boundary, and the bulk metric is varied by increasing the mass of the black hole. In this section we put both these proposals to test. Namely, we numerically investigate the behavior of holographic quantities for the CFT dual to the charged AdS black hole spacetimes.  We start with the entanglement entropy.

\subsection{Entanglement entropy}

For any quantum system localized to some region $A$, the entanglement entropy is given by
\begin{equation}\label{entShan}
S_{A}= - \text{Tr}_B \rho_{A} \log \rho_{A} \,,
\end{equation}
where the system is partitioned into region $A$ and its complement $B$ where $ \rho_{A}=\text{Tr}_{B}\qket{\psi}\qbra{\psi}$ is the reduced density matrix describing subsystem $A$ with the system originally being in a pure state
$\qket{\psi}$. A common example is that of complementary spatial volumes  on a given constant time slice, their common boundary being the  ``entangling surface''.   One can express $\rho_{A}$ in the form of an effective thermal system
\begin{equation} \label{eqn:mod}
\rho_{A}= \frac{e^{-H_{A}/T_{0}}}{\text{Tr}(e^{-H_{A}/T_{0}})}\, ,
\end{equation}
where $H_{A}$ is known as the {\it modular Hamiltonian}, {and $T_0$ is a constant with units of temperature}.  Upon employing \eqref{entShan} this yields the first law
\begin{equation}
\label{eqn:ee}
T_{0} dS_{A} = \text{Tr}\bigl(H_{A} d\rho_{A}\bigr) \equiv d \left< H_{A}\right>\,
\end{equation}
for  entanglement entropy  \cite{Blanco:2013joa,Wong:2013gua}.

The Ryu--Takayanagi proposal  \cite{Ryu:2006bv} extends the above construction to that of a
 CFT in $d-1$ dimensions constructed in a spacetime corresponding to the boundary of an asymptotically bulk AdS{$_d$} spacetime (for which the quantum state of the CFT  is not necessarily pure).  Continuing to refer to $S_{A}$
 as the entanglement entropy, their proposal states that
 \begin{equation} \label{eqn:ryu}
S_{A} = \frac{A_{\Sigma}}{4G_d}
\end{equation}
applied to a bulk minimal surface $\Sigma$ (with area $A_{\Sigma}$),
whose boundary matches the entangling surface $A$ in the CFT at spatial infinity.  To compute this quantity a regularization procedure is required since the minimal surface area in an asymptotically AdS bulk is formally divergent.
 In what follows, rather than the entanglement entropy of the excited CFT state (in the presence of a black hole), we are interested in the {\em relative entanglement entropy}
\be\label{relative}
S_E=S_A-S^{(0)}_A\,,
\ee
given by subtracting the analogous contribution $S^{(0)}_A$ from vacuum AdS.

Let us turn now to the calculation of the relative entanglement entropy in the
charged AdS black hole spacetime.
Choosing the region $A$ to be a spherical cap (as in \cite{Nguyen:2015wfa}), the entangling surface can then be described by constant polar angle $\theta=\theta_0$, and   the entanglement entropy obtained via \eqref{eqn:ryu}, where
the area $A_{\Sigma}$ is obtained by minimizing the action functional
\be\label{entanglement_entropy_action}
A_{\Sigma}=\omega_{ (d-3)}\int_0^{\theta_0} (r(\theta)\sin\theta)^{d-3}\sqrt{\frac{r'(\theta)^2}{f(r(\theta))}+r(\theta)^2}\,d\theta
\ee
via Euler-Lagrangian variation.
The relative entanglement entropy \eqref{relative}
is given by subtracting the analogous contribution from vacuum AdS. This latter contribution is explicitly known \cite{Johnson:2013dka}:
\be\label{r_AdS}
r_{0}(\theta)=l\lb \lb\frac{\cos\theta}{\cos\theta_0}\rb^2-1\rb^{-\frac{1}{2}}
\ee
and the corresponding quantity $S_A^{(0)}$ straightforwardly computed. However the Euler--Lagrange system following from
\eqref{entanglement_entropy_action} must in general be solved numerically  with  boundary conditions
\begin{gather}
 r(\theta_0)\rightarrow\infty,\label{cond_1}\\
 r'(0)=0.\label{cond_2}
 \end{gather}
where (\ref{cond_1}) ensures that $r(\theta)$ coincides with the entangling surface on the boundary $r\rightarrow \infty$ and (\ref{cond_2}) ensures regularity at the centre ($\theta=0$, the middle of the entangling surface, which is the point of maximum penetration into the bulk).   Since entanglement entropy is divergent, a long-distance cut off must be introduced for regularization, which can be implemented by choosing a cut-off value  $\theta_c<\theta_0$, and only integrating up to $\theta_c$.
In our investigation we limit ourselves to considering small $\theta_0$.

\begin{figure}[h!]
\begin{center}
\includegraphics[scale=0.55]{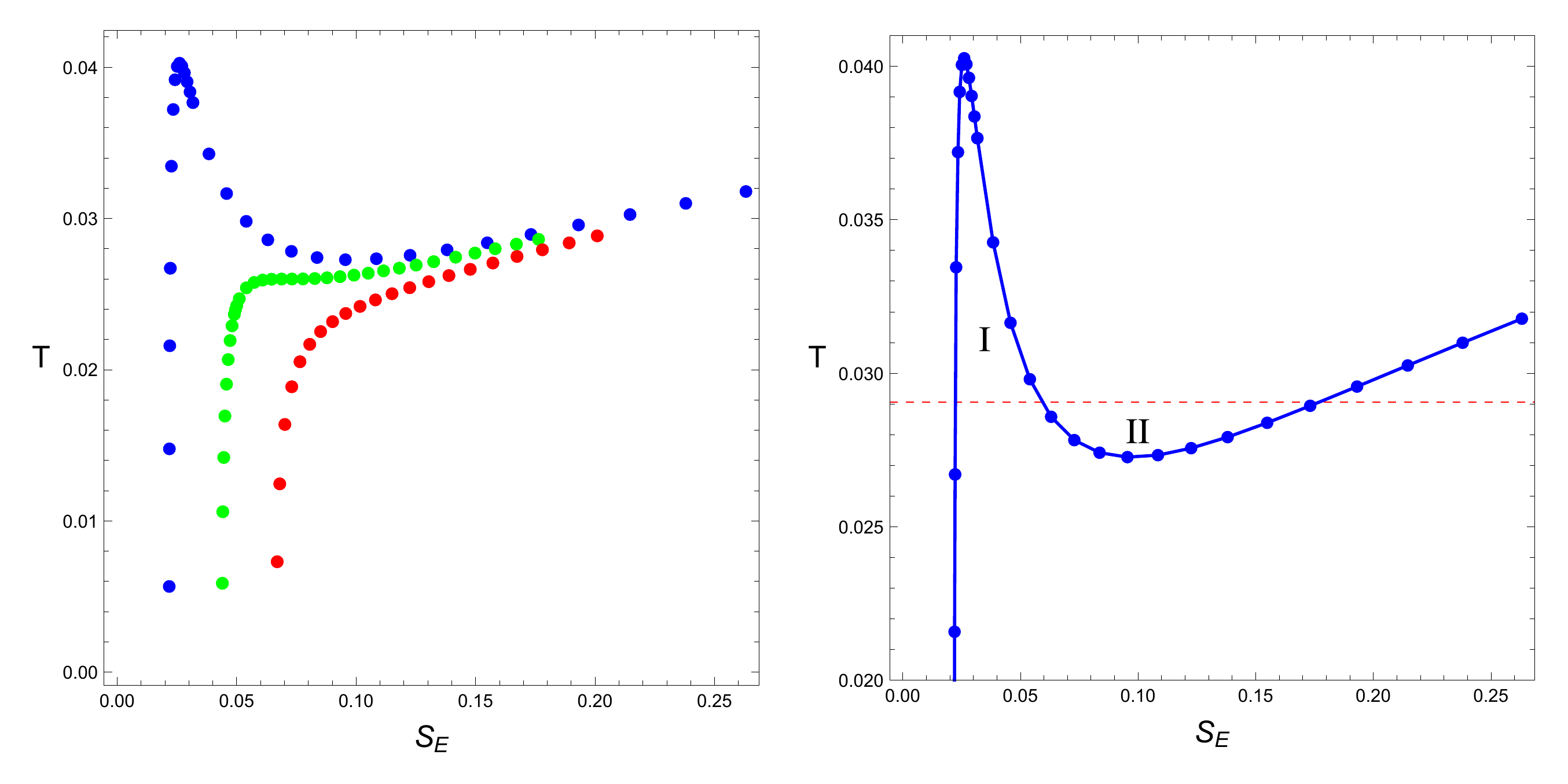}
\end{center}
\caption{ \textbf{
$T-S_E$ diagram for a $d=4$ charged AdS black hole.}
On the left we see the oscillatory behavior for $Q<Q_{crit}$ (blue) and the point of inflection for $Q=Q_{crit}$ (green). On the right the Maxwell construction for entanglement entropy for $Q<Q_{crit}$ is shown; it was claimed in \cite{Nguyen:2015wfa} that Areas I and II as shown are equal above and below the phase transition temperature of the black hole $T_*$, given by the solid line. {The paramaters chosen were $l=10$, $\theta_0=0.15$, and $\theta_c=0.149$. Again, the values of $Q$ are $Q=0.5\, Q_{crit}$ (blue), $Q=Q_{crit}$ (green) and $Q=1.5\, Q_{crit}$ (red).}}
\label{fig:entanglement_behavior}
\end{figure}

After computing $S_E$ for a range of values of $T$ for a charged AdS black hole in 3+1 dimensions, we can plot the isocharges in the $T$/$S_E$ plane; see Fig. \ref{fig:entanglement_behavior}. Comparing Fig.~\ref{fig:phase_behavior} and  Fig.~\ref{fig:entanglement_behavior}, which show the isocharges in the $T$/$S$ and $T$/$S_E$ planes respectively, we see that
 the behavior of black hole temperature against entanglement entropy is {\em qualitatively similar} to that against the black hole entropy. In particular, we see oscillatory behavior for charges below the critical charge $Q_{crit}$, a point of inflection at $Q_{crit}$, and monotonic increase above $Q_{crit}$.  It is perhaps natural to consider that an equal area law holds for $Q<Q_{crit}$ (such that  Areas I and II are equal on the right-side of
 Fig. \ref{fig:entanglement_behavior}), and indeed numerical evidence in favour of this has been presented \cite{Nguyen:2015wfa}.

\begin{figure}[h!]
\begin{center}
\includegraphics[scale=0.4]{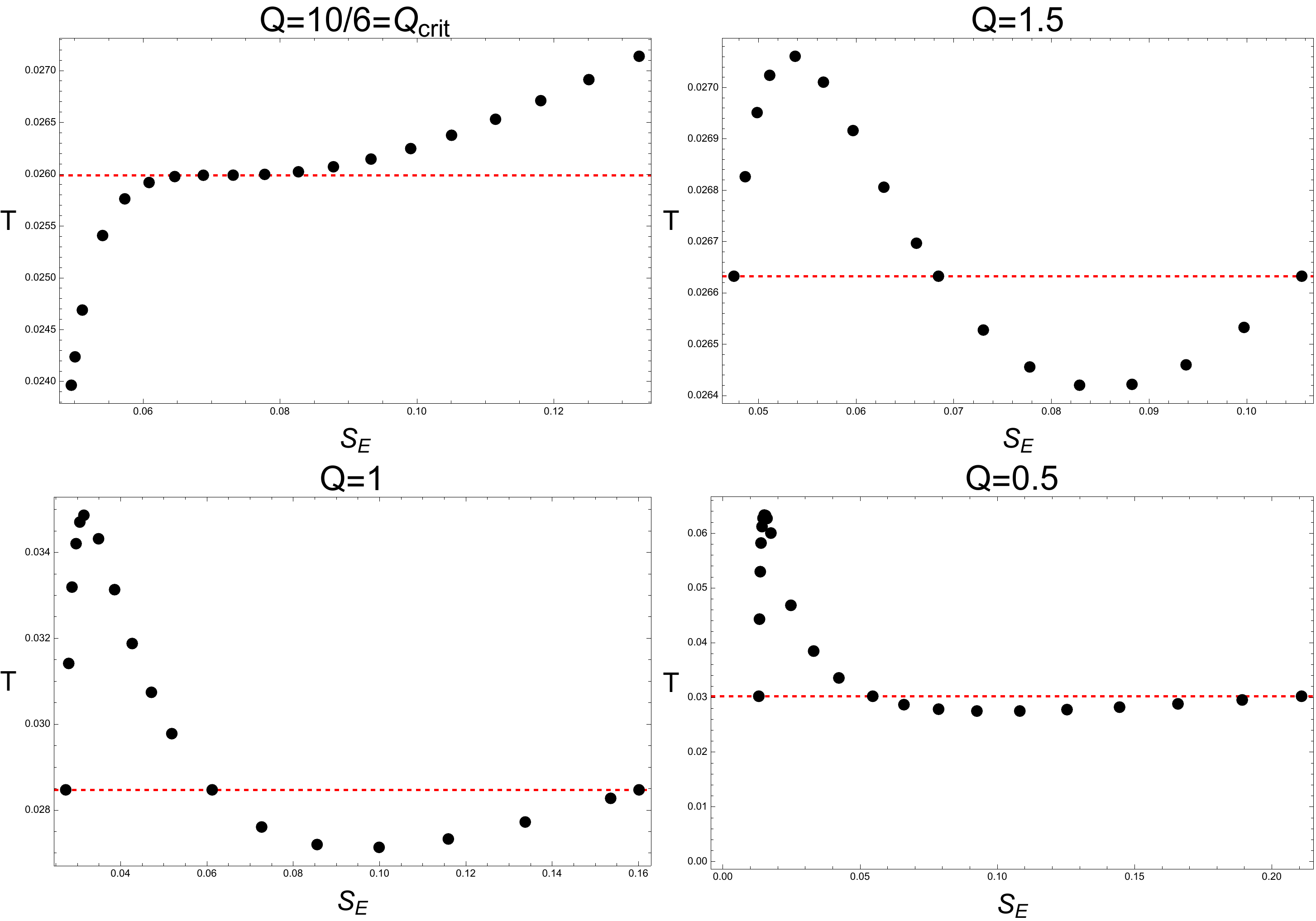}
\caption{\textbf{Charge dependence of $T-S_{E}$ diagrams.} The plots are displayed for the $d=4$ charged ADS black holes, with $l=10$, $\theta_0=0.15$, and $\theta_c=0.149$. The phase transition temperature $T_*$ is plotted in red in each case.}
\label{fig:4d_ee}
\end{center}
\end{figure}

 However, as noted in \cite{Sun:2016til} the apparent validity of the equal area law seems misleading, with
the discrepancy growing as isocharges are chosen further away from near critical values, originally considered in
\cite{Nguyen:2015wfa}. To resolve this dispute, we have displayed in Fig.~\ref{fig:4d_ee} the
$T$ versus $S_E$ diagrams for various isocharges.
The phase transition temperature $T_*$ is identified as the phase transition temperature in the bulk, where the isocharge intersects itself in the free energy diagram in Fig. \ref{fig:phase_behavior}. The results of the equal area law are shown in Table~\ref{table:equal_area_results_4d}, with the relative error defined as $\frac{\mathrm{Area(I)}-\mathrm{Area(II)}}{\mathrm{Area(I)}} \times 100$. We see that at values of $Q$ very close to criticality, the relative error is low enough to lead one to believe an equal area law might hold; however, as we move away from criticality, the equal area law breaks down, as noted in \cite{Sun:2016til}.
We checked that the equal area law also breaks down for charged AdS black holes in higher spacetime dimensions.

\begin{table}[h!]
\begin{center}
\begin{tabular}{|c|c|c|c|c|c|}
\hline$Q$&$Q/Q_c$&$T_*$&Area(I)&Area(II)&Relative error\\
\hline\hline
1.5&0.9&0.0266324&$5.309\times10^{-6}$&$5.202\times10^{-6}$&2.02\%\\
\hline
1&0.6&0.02847&$1.040\times 10^{-4}$&$8.425\times10^{-5}$&19.0\%\\
\hline
0.5&0.3&0.030198&$4.745\times 10^{-4}$&$2.653\times 10^{-4}$&44.1\%\\
\hline
\end{tabular}
\caption{
\textbf{Failure of the equal area law in the $T-S_E$ plane in $d=4$.} In this table $l=10$ and $Q_c=5/3$. We see that as we move further from criticality, the relative error between Areas I and II increases and the equal area law does not hold. The plots of the results can be found in Fig. \ref{fig:4d_ee}.}
\label{table:equal_area_results_4d}
\end{center}
\end{table}

 The disagreement between our findings and those in \cite{Nguyen:2015wfa, Zeng:2015tfj, Liu:2017jbm,Zeng:2015wtt, Zeng:2016sei, Mo:2016cmi,Mo:2016ijb, Zeng:2016aly, Zeng:2016fsb, ElMoumni:2016eqh, Li:2017gyc, Li:2017xiv, Zeng:2017zlm} arises both from the fact that we have probed further away from criticality than had been done, and are thus finding larger relative errors, and also from a numerical argument. As noted in \cite{ Sun:2016til}, when the areas given by (\ref{maxwell_larger_areas}) are compared, there is less of a relative error than for the more precise areas given by (\ref{integration_areas}).
As we shall elaborate on below, such discrepancies can be huge; as much as  3 and 42 percent for $Q=0.3\,Q_c$.

\subsection{Two-point correlation function}

We turn now to the equal-time two-point correlation function \cite{Balasubramanian:1999zv}
 \be
 \left<\mathcal{O}(t_0,x_i)\mathcal{O}(t_0,x_j)\right>\approx e^{-\Delta L\lb x_i,x_j\rb},
 \ee
 where $L\lb x_i, x_j \rb $ is the smallest bulk geodesic between $(t_0,x_i)$ and $(t_0,x_j)$.
 In order to formulate a Maxwell construction for the two-point correlation function, we can choose the points $x_1=\lb\theta=0,\phi=\frac{\pi}{2}\rb$, $x_2=\lb\theta=\theta_0,\phi=\frac{\pi}{2}\rb$. $L\lb x_1, x_2 \rb $ can be then computed by minimising the functional
 \be
 L\lb x_1, x_2 \rb =\int_0^{\theta_0}\sqrt{\frac{r'(\theta)^2}{f(r(\theta))}+r(\theta)^2}d\theta\,.
 \ee
 This is a similar computation to that carried out using \eqref{entanglement_entropy_action}. The quantity $L$ must be computed by solving the Euler--Lagrange equations, a cut off $\theta_c$ is chosen, and the vacuum AdS two point function $L_0$ is be subtracted off to obtain $\Delta L=L-L_0$. Again, the vacuum AdS solution is given by equation (\ref{r_AdS}).

\begin{figure}[h!]
\begin{center}
\includegraphics[scale=0.4]{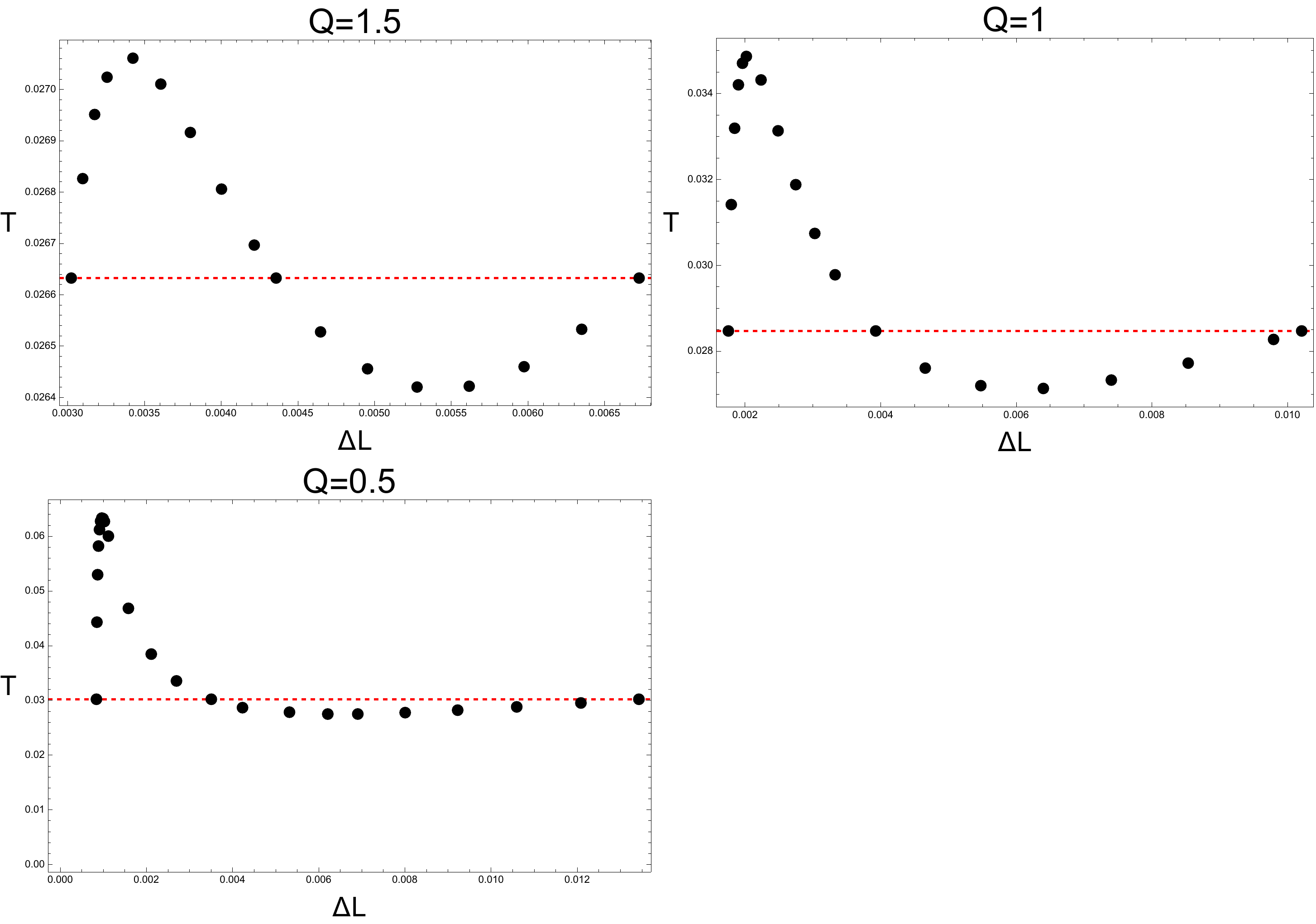}
\caption{\textbf{Charge dependence of $T-\Delta L$ diagrams.}
The plots are displayed for the $d=4$ charged ADS black holes. The phase transition temperature $T_*$ is plotted in red in each case. Again we have {$l=10$, $\theta_0=0.15$, and $\theta_c=0.149$.} }
\label{fig:4d_cf}
\end{center}
\end{figure}

 Contrary to claims \cite{Zeng:2015wtt, Zeng:2016sei, Mo:2016cmi,Mo:2016ijb, Zeng:2016aly, Zeng:2016fsb, ElMoumni:2016eqh, Li:2017gyc, Li:2017xiv, Zeng:2017zlm} that when computed in this way $L$ obeys a Maxwell equal area construction, we find again that there is no equal area law in this plane as we move away from criticality. Our results are illustrated in  Fig. \ref{fig:4d_cf} and Table \ref{table:equal_area_results_correlation_function}, where the relative error between Areas I and II is $45\%$ at $Q=0.3\, Q_{crit}$.

\begin{table}[h!]
\begin{center}
\begin{tabular}{|c|c|c|c|c|c|}
\hline$Q$&$Q/Q_c$&$T_*$&Area(I)&Area(II)&Relative error\\
\hline\hline
1.5&0.9&0.0266324&$3.382\times10^{-7}$&$3.313\times10^{-7}$&2.01\%\\
\hline
1&0.6&0.02847&$6.685\times 10^{-6}$&$5.350\times10^{-6}$&20.0\%\\
\hline
0.5&0.3&0.030198&$3.032\times 10^{-5}$&$1.667\times 10^{-5}$&45.0\%\\
\hline
\end{tabular}
\caption{\textbf{The failure of the equal area law for the two-point correlation function.} Similarly as for $S_E$, as we move away from criticality we see a failure in the equal area law for $\Delta L$.}
\label{table:equal_area_results_correlation_function}
\end{center}
\end{table}

\section{Failure of the equal area construction}

 Our results show that there is no  numerical evidence for a holographic equal area law in either the black hole temperature/entanglement entropy or the black hole temperature/two point correlation function plane, at the phase transition temperature of the black hole.

Similar evidence \cite{Sun:2016til} for the failure of the equal area law for entanglement entropy  has been attributed to the first law of entanglement \eqref{eqn:ee}, which we rewrite as \cite{Bhattacharya:2012mi}
 \be
 dE_A=T_{ent} dS_A,
 \ee
where $E_A$ is the energy contained in a region $A$ and $S_A$ is the entanglement entropy between  a small region $A$ and its complement. We note that our restriction to only small values of $\theta_0$ ensures our calculations are well within the small-region regime.   $T_{ent}$ is known as the entanglement temperature,  defined by comparing the energy to entropy ratio $\frac{\Delta E_A}{\Delta S_A}$ for an excited state of the region A relative to the ground state for the same region in the CFT.

\begin{figure}[h!]
\begin{center}
\includegraphics[scale=0.4]{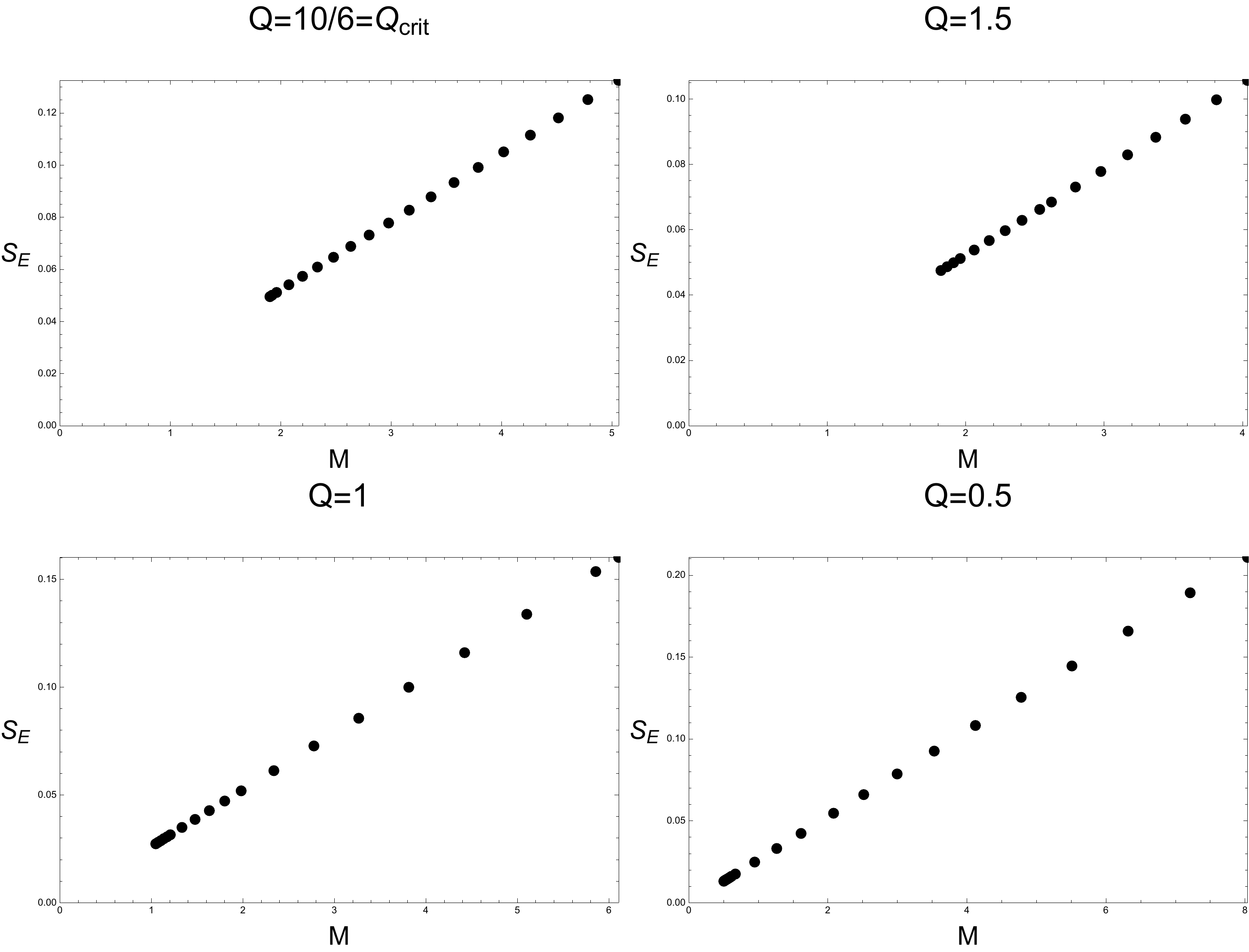}
\caption{\textbf{ The first law for relative entanglement entropy.} These plots show various isocharges in the black hole mass-entanglement entropy plane for the charged AdS black hole in $d=4$ dimensions. The graphs verify that $S_E\propto M$ and thus verify the first law of entanglement entropy.
}\label{fig:mass_ee}
\end{center}

\end{figure}

$E_A$ can be computed by integrating the stress tensor of the black hole spacetime
on the boundary \tcb{\cite{PhysRevD.47.1407}}:
 \be
 E_A=\int d^{d-2}x\,T_{tt}.
 \ee
 For spherically symmetric asymptotically AdS spacetime, upon computing the difference
 $\Delta E_A \equiv E_A-E^{(0)}_A$ between the excited and vacuum state in region A
  this becomes
 \be
 \Delta E_A\propto\int d^{d-2}x M,\label{deltaE_mass}
 \ee
 since
 $T_{tt}$ is proportional to the mass $M$ \cite{Balasubramanian:1999re}.
  For constant entangling region size we should therefore have
 \be
 T_{ent} S_E\propto M
 \ee
using \eqref{relative};
since $T_{ent}$ depends only on entangling region size \cite{Bhattacharya:2012mi}, which is kept constant, we find
 \be
S_E\propto M
 \ee
 as a {\em first law for the relative entanglement entropy}. This relation is straightforwardly tested numerically.
 Indeed, when we plot $S_E$ against $M$ we find that these are proportional, as depicted in  Fig. \ref{fig:mass_ee}.  From these graphs we see that while \cite{Bhattacharya:2012mi} only dealt with uncharged, asymptotically planar AdS spacetimes, its results are also valid in the charged asymptotically spherical case.

 \begin{figure}[h!]

\begin{center}
\includegraphics[scale=0.4]{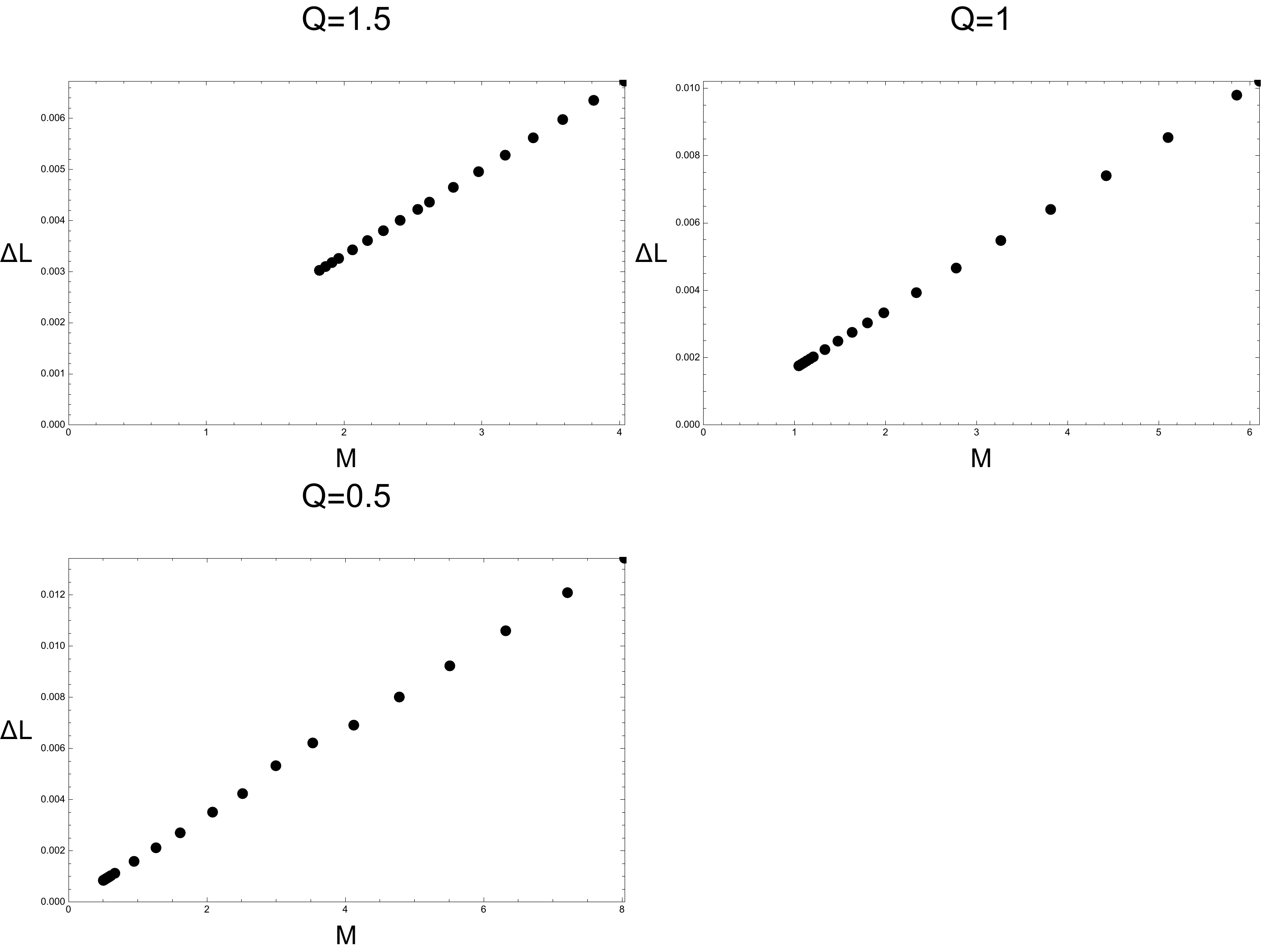} 	
\caption{ \textbf{ The proportionality of $\Delta L$ and $M$.} These plots show the isocharges in the black hole mass-$\Delta L$ plane for the charged AdS black hole in $d=4$ dimensions. The graphs verify that $\Delta L\propto  M$ and thus that there is not an equal area law for $\Delta L$}
\label{fig:mass_cf}
\end{center}
\end{figure}

 Thus, defining for an equal area law for entanglement entropy is equivalent to defining an equal area law in the $T$/$M$ plane. However, the Maxwell construction works in general only for pairs of conjugate thermodynamical variables: there is no equal area law in the $T$/$M$ plane.
Thus, there is no reason to expect an equal area law in the $T$/$S_E$ plane. To study an equal area law for entanglement entropy,
we would require a corresponding thermodynamic interpretation of the free energy, and we would need to consider $S_E$ plotted against its thermodynamic conjugate.
We likewise find that the two-point correlation function is proportional to $M$, as shown in  Fig. \ref{fig:mass_cf}.

As noted previously, one of the main reasons for the discrepancy between our findings and those
contending an equal area law has to do with using
 (\ref{maxwell_larger_areas}) instead of the more precise (\ref{integration_areas}). In Table \ref{table:two_areas_comparison}  we illustrate this for several values of $Q/Q_c$.  It is clear that
the distinction can be very large, and it is clear that (\ref{integration_areas}) provides no support for
an equal area law.

\begin{table}[h!]
\begin{center}
\begin{tabular}{|c|c|c|}
\hline$Q/Q_c$&\multicolumn{2}{c|}{Relative Error}\\
\cline{2-3}
&Areas: Eq. (\ref{maxwell_larger_areas})&Areas: Eq. (\ref{integration_areas})\\
\hline\hline
0.9&0.00389\%&1.12\%\\
\hline
0.6&0.455\%&16.7\%\\
\hline
0.3&3.17\%&42.1\%\\
\hline
\end{tabular}
\caption[Caption for LOF]{ \textbf{Comparing the Accuracy of the errors given by Equations (\ref{maxwell_larger_areas}) and (\ref{integration_areas}).} We have computed the relative error on the $T$/$M$ plane between the areas defined by each equation. Due to the areas being defined by (\ref{maxwell_larger_areas}) being much larger, their relative error is less accurate than between the relative error between the areas of interest defined by (\ref{integration_areas}).\footnotemark}
\label{table:two_areas_comparison}
\end{center}
\end{table}

We summarize in Table \ref{table:comparison_of_all_planes} a comparison of the relative errors between Areas I and II on the $T$/$M$, $T$/$S_E$, and $T$/$\Delta L$ planes. In all cases we find that this
quantity grows as the departure from criticality increases.  We conclude that there no reason to expect an equal area law for either the two-point correlation function or for the entanglement entropy.
\begin{table}[h!]
\begin{center}
\begin{tabular}{|c|c|c|c|c|}
\hline$Q$&$Q/Q_c$&\multicolumn{3}{c|}{Relative Error}\\
\cline{3-5}
&&Mass&HEE&$\Delta L$\\
\hline\hline
1.5&0.9&1.96\%&2.02\%&2.01\%\\
\hline
1&0.6&18.8\%&19.0\%&20.0\%\\
\hline
0.5&0.3&44.2\%&44.1\%&45.0\%\\
\hline
\end{tabular}
\caption[Caption for LOF]{ \textbf{Comparing the Relative Errors.} We have computed the relative error between Areas I and II on the Temperature/Mass plane and we have compared this with those on the $T$/$\Delta S$ and $T$/$\Delta L$ planes. The numerical errors in both cases are very close to that in the $T$/$M$ plane.
}
\label{table:comparison_of_all_planes}
\end{center}
\end{table}
\color{black}

We close this section by commenting on the $\theta_c$ and $\theta_0$ dependence of our results in $S_E=S_A-S_A^{(0)}$. This manifests itself differently in $S_A$ and $S_A^{(0)}$ in such a way that the $\theta_c$ dependence of $S_E$ was numerically found to be given by the following expression\footnotetext{ The discrepancies between the values given for the relative errors in the $T$/$M$ plane between Tables \ref{table:two_areas_comparison} and \ref{table:comparison_of_all_planes} comes from the fact that the areas in table \ref{table:two_areas_comparison} were found exactly, whereas the areas in table \ref{table:comparison_of_all_planes} were numerically integrated using only the masses at the points for which the $S_E$ and $\Delta L$ values were calculated, in order to obtain a more meaningful comparison.}
\be
S_E=\sin\theta_0\int_0^{\theta_c} \sin^{d-3}x\, dx\,\,F\lb Q,r_+,l\rb.
\ee
These relationships are not found in $S_A$ or in the background entanglement entropy $S_A^{(0)}$, but only in their difference $S_E$. We have numerically checked this for $d=4$ and $5$. Thus the proportionality between $S_E$ and $M$ is only dependent on either of these parameters via the proportionality constant: the slopes of the lines in Figs \ref{fig:mass_ee} and \ref{fig:mass_cf} will change, but the relative error between Areas I and II is uneffected.

\section{The ``approximate'' equal area law near criticality}
\label{sec:approximate}

It is evident that much of the confusion in the literature on the subject of the holographic equal area law stems from the seemingly ``approximate'' equal area law obeyed on the $T/M$ plane near criticality. In 4 spacetime dimensions this can be explained by demonstrating that Areas I and II on the $T/M$ plane must approach zero at the same rate near criticality, which we can see by Taylor expanding expressions for these areas near criticality.

Namely, in 4 spacetime dimensions an expression for the phase transition temperature  $T_*$, obtained by requiring that both the temperature and the free energy are equal for the large and small black holes,
is exactly known \cite{Mo:2016sel, Kubiznak:2016qmn}
\be
T_*=\frac{\sqrt{l-2Q}}{l^{3/2}\pi}\,.
\ee
Areas I and II are given by
\begin{align}
\text{Area(I)}=\int _{M_1}^{M_2}TdM-T_*(M_2-M_1)\,,\\
\text{Area(II)}=T_*(M_3-M_2)-\int _{M_2}^{M_3}TdM\,,
\end{align}
where $M_1$, $M_2$, and $M_3$ are the masses $M(S_i,Q)$ corresponding to the three solutions $S_1$, $S_2$, $S_3$ of $T(S,Q)=T_*$. We note that the Areas (I) and (II) as defined above approaching zero 
at the same rate is equivalent to the larger areas $\text{Area(A)}=T_*(M_3-M_1)$ and $\text{Area(B)}=\int_{M_1}^{M_3}TdM$ approaching zero 
at the same rate and so it suffices to look at the near-critical expansions of areas A and B. $S_1$ and $S_3$ are known \cite{Mo:2016sel, Kubiznak:2016qmn}:
\begin{align}
S_1=\frac{4 l^2\pi Q^2}{\left(\sqrt{l(l-6 Q)}+\sqrt{l(l-2Q)}\right)^2}\,,\quad
S_3=\frac{\pi}{4}\left(\sqrt{l(l-6 Q)}+\sqrt{l(l-2Q)}\right)^2\,.
\end{align}
This allows us to find expressions for $M_1$ and $M_3$ from $M=\frac{S^2+l^2\pi(\pi Q^2+S)}{2l^2\pi^{3/2}\sqrt{S}}.$ After doing this, we can expand areas A and B about the critical charge $Q_{crit}=l/6$, with $\delta q=(Q_c-Q)/l$:
\ba
\text{Area(A)}&=&T_*(M_3-M_1)\nonumber\\
&=&\frac{4}{3\pi}\left(\delta q\right)^{1/2}+\frac{6}{\pi}\left(\delta q\right)^{3/2}+\frac{9}{2\pi}\left(\delta q\right)^{5/2}-
\frac{9}{4\pi}\left(\delta q\right)^{7/2}+\dots.\label{areaaexp}\\
\text{Area(B)}&=&\int_{M_1}^{M_3}T(S,Q)dM=\int_{S_1}^{S_3}T\frac{\partial M}{\partial S}dS\nonumber\\
&=& \frac{4}{3\pi}\left(\delta q\right)^{1/2}+\frac{6}{\pi}\left(\delta q\right)^{3/2}+\frac{9}{2\pi}\left(\delta q\right)^{5/2}+\frac{2277}{140\pi}\left(\delta q\right)^{7/2}+\dots\,,\label{areaexpansion}
\ea
from which we can see that Areas A and B agree up to the first three terms of the Taylor expansion. This explains the appearance of an equal law near criticality when  $\delta q$ is small.
A slightly more general argument, valid for other black holes and in any dimension, is presented in Appendix~\ref{AppA}.


\section{Conclusions}

The observation \cite{Johnson:2013dka} that  relative entanglement entropy displays qualitatively similar behaviour to black hole entropy on isocharges of the $3+1$-dimensional charged AdS black hole below criticality seemed intriguing. Nguyen attempted to sharpen this similarity in \cite{Nguyen:2015wfa} by showing that the  relative entanglement entropy obeys an equal area construction, evidence for a phase transition. Further claims that a holographic equal area law holds in the $T$/geodesic length plane have also been put forward  (see references above).

\begin{figure}[h!]
\begin{center}
\includegraphics[scale=0.6]{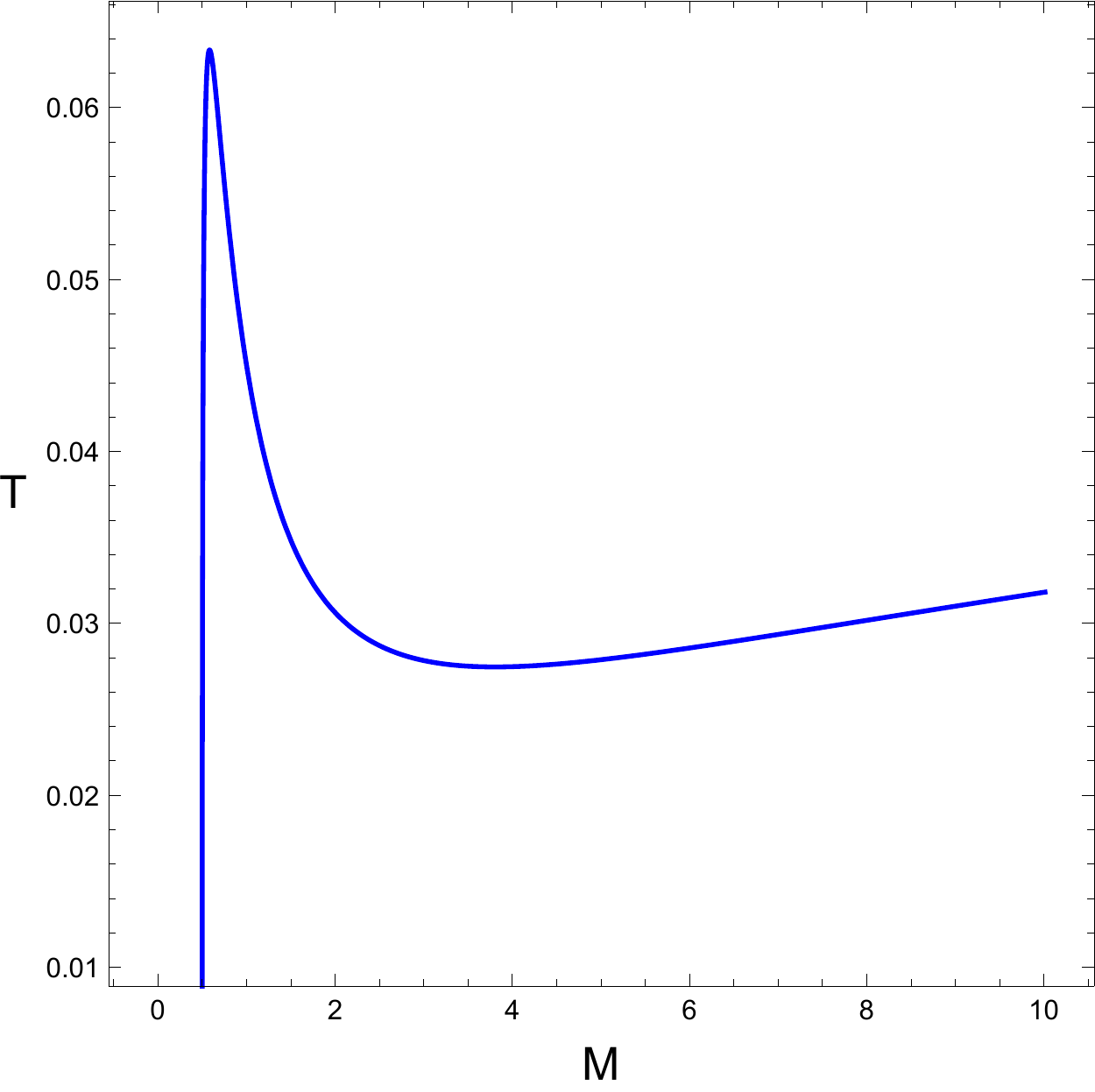}
\end{center}
\caption{ {\bf $T$ versus $M$.} $T$ is plotted against $M$ in $d=4$, with $l=10$ and $Q=0.5=0.3\,Q_c$. We see the oscillatory behaviour of $T$; our $T$/$S_E$ and $T$/$\Delta L$ graphs (Figs \ref{fig:4d_ee} and \ref{fig:4d_cf}) are just rescaled versions of this plot.}
\label{fig:t_v_m}
\end{figure}

Our results indicate, commensurate with \cite{Sun:2016til},  that all proposals thus far put forward that
 a form of Maxwell's equal area law holds for entanglement entropy are false {(although it is `almost satisfied' near criticality, see Table~\ref{table:comparison_of_all_planes})}.  Moreover, we do not find it surprising that the isocharges display oscillatory behaviour below $Q_{crit}$ and a point of inflection at $Q_{crit}$ on the $T-S_E$ plane.
 This is a simple consequence of two facts: i) $S_E=S_E(M)$ is a monotonic function due to the first law for the relative entanglement entropy and ii)
 temperature $T$, when displayed as a function of $M$,  demonstrates oscillatory
 behavior, as shown in  Fig.~\ref{fig:t_v_m}, with the oscillation disappearing at criticality. 

After numerically studying the holographic Maxwell construction for entanglement entropy and the two-point correlation function, we find no reason to support such proposals for the equal area law. The entanglement entropy is not dual to the black hole entropy, and it should not be expected that it obeys an equal area law.
Any equal area construction for entanglement entropy should be studied in relation to its thermodynamic dual,
and  any claim of a phase transition must be backed up by a free energy diagram on the boundary similar to that in Fig. \ref{fig:phase_behavior}; in other words there must be an analogue of free energy that displays swallowtail behavior.

While we expect that phase transition for a bulk black hole has a counterpart in the boundary CFT,  although the entanglement entropy jumps in such a phase transition, the transition temperature is not given by the associated equal area law.
A CFT phase transition will be governed by the corresponding free energy of CFT.  Translating this into a holographic equal area law of
some kind remains an open question.

\section*{Acknowledgments}
{We would like to thank the anonymous referee for helping us to improve our manuscript.}
This research was supported in part by Perimeter Institute for Theoretical Physics and by the Natural Sciences and Engineering Research Council of Canada. Research at Perimeter Institute is supported by the Government of
Canada through the Department of Innovation, Science and Economic
Development Canada and by the Province of Ontario through the
Ministry of Research, Innovation and Science.

\appendix

\section{The approximate equal area law: general argument}\label{AppA}

In section \ref{sec:approximate} we presented an argument as to why 
there appears to be an ``approximate'' equal area law near criticality in the $T/M$ plane for charged AdS black holes in 4 dimensions. Here we demonstrate that this holds more generally,
in any number of dimensions and for other black hole solutions for which the expressions for $T_*$, and $M_1$ and $M_3$  are not explicitly known. 

Areas A and B are given by
\begin{align}
\text{Area(A)}&=T_*(M_3-M_1)\,,\\
\text{Area(B)}&=\int_{M_1}^{M_3}TdM\,.\label{Area_b}
\end{align}
Assuming that for a chosen black hole there 
is an equal area law satisfied in the $T$/$S$ plane, we can find an expression for $T_*$:
\be
T_*=\frac{\int_{S_1}^{S_3}TdS}{S_3-S_1}\,,
\ee
such that
\be
\text{Area(A)}=\frac{M_3-M_1}{S_3-S_1}\int_{S_1}^{S_3} TdS\,.\label{Area_a}
\ee

To calculate the integrals \eqref{Area_b} and \eqref{Area_a} we expand
\ba
T=T(S,Q)&=&T_c+\frac{\partial T}{\partial S}\Delta S+\frac{\partial T}{\partial Q}\Delta Q+\dots\nonumber\\
&=&T_c+\frac{\partial T}{\partial Q}\Delta Q+\dots\,,
\ea
where $\Delta Q\equiv Q-Q_c$, $\Delta S=S-S_c$, and it is understood that the derivatives are evaluated at $Q_c, S_c$; the last equality follows from the fact that
that at criticality $\partial T/\partial S=0$. So we have
\ba\label{Aa}
\text{Area(A)}&=& \frac{M_3-M_1}{S_3-S_1}\int_{S_1}^{S_3}\Bigl(T_c+\frac{\partial T}{\partial Q}\Delta Q\Bigr)dS\nn\\
&=&
\Bigl(T_c+\frac{\partial T}{\partial Q}\Delta Q\Bigr)(M_3-M_1)\,.\label{ar_a}
\ea
At the same time we have
\ba\label{Bb}
\text{Area(B)}&=&\int _{M_1}^{M_3}TdM=\int_{M_1}^{M_3}\Bigl(T_c+\frac{\partial T}{\partial Q}\Delta Q\Bigr)dM\nonumber\\
&=&
\Bigl(T_c+\frac{\partial T}{\partial Q}\Delta Q\Bigr)(M_3-M_1)\,,
\ea
and hence the areas are equal to this order of expansion in $\Delta Q$. Since $(M_3-M_1)$ has to go to zero near criticality, the above formulas show that the areas are equal at least to the order linear in $\Delta Q$.

\bibliography{references}

\providecommand{\href}[2]{#2}\begingroup\raggedright\begin{thebibliography}{10}

\bibitem{Spallucci:2013osa}
E.~Spallucci and A.~Smailagic, \emph{{Maxwell's equal area law for charged
  Anti-deSitter black holes}},
  \href{https://doi.org/10.1016/j.physletb.2013.05.038}{\emph{Phys. Lett.}
  {\bfseries B723} (2013) 436--441},
  [\href{https://arxiv.org/abs/1305.3379}{{\ttfamily 1305.3379}}].

\bibitem{Lan:2015bia}
S.-Q. Lan, J.-X. Mo and W.-B. Liu, \emph{{A note on Maxwell's equal area law
  for black hole phase transition}},
  \href{https://doi.org/10.1140/epjc/s10052-015-3641-0}{\emph{Eur. Phys. J.}
  {\bfseries C75} (2015) 419},
  [\href{https://arxiv.org/abs/1503.07658}{{\ttfamily 1503.07658}}].

\bibitem{Xu:2015hba}
H.~Xu and Z.-M. Xu, \emph{{Maxwell's equal area law for Lovelock
  Thermodynamics}}, \href{https://doi.org/10.1142/S0218271817500377}{\emph{Int.
  J. Mod. Phys.} {\bfseries D26} (2017) 1750037},
  [\href{https://arxiv.org/abs/1510.06557}{{\ttfamily 1510.06557}}].

\bibitem{Kubiznak:2016qmn}
D.~Kubiznak, R.~B. Mann and M.~Teo, \emph{{Black hole chemistry: thermodynamics
  with Lambda}}, \href{https://doi.org/10.1088/1361-6382/aa5c69}{\emph{Class.
  Quant. Grav.} {\bfseries 34} (2017) 063001},
  [\href{https://arxiv.org/abs/1608.06147}{{\ttfamily 1608.06147}}].

\bibitem{Nguyen:2015wfa}
P.~H. Nguyen, \emph{{An equal area law for holographic entanglement entropy of
  the AdS-RN black hole}},
  \href{https://doi.org/10.1007/JHEP12(2015)139}{\emph{JHEP} {\bfseries 12}
  (2015) 139}, [\href{https://arxiv.org/abs/1508.01955}{{\ttfamily
  1508.01955}}].

\bibitem{Zeng:2015tfj}
X.-X. Zeng, H.~Zhang and L.-F. Li, \emph{{Phase transition of holographic
  entanglement entropy in massive gravity}},
  \href{https://doi.org/10.1016/j.physletb.2016.03.013}{\emph{Phys. Lett.}
  {\bfseries B756} (2016) 170--179},
  [\href{https://arxiv.org/abs/1511.00383}{{\ttfamily 1511.00383}}].

\bibitem{Sun:2016til}
Y.~Sun, H.~Xu and L.~Zhao, \emph{{Thermodynamics and holographic entanglement
  entropy for spherical black holes in 5D Gauss-Bonnet gravity}},
  \href{https://doi.org/10.1007/JHEP09(2016)060}{\emph{JHEP} {\bfseries 09}
  (2016) 060}, [\href{https://arxiv.org/abs/1606.06531}{{\ttfamily
  1606.06531}}].

\bibitem{Liu:2017jbm}
X.-M. Liu, H.-B. Shao and X.-X. Zeng, \emph{{Van der Waals-like phase
  transition from holographic entanglement entropy in Lorentz breaking massive
  gravity}},  \href{https://arxiv.org/abs/1706.04431}{{\ttfamily 1706.04431}}.

\bibitem{Zeng:2015wtt}
X.-X. Zeng and L.-F. Li, \emph{{Van der Waals phase transition in the framework
  of holography}},
  \href{https://doi.org/10.1016/j.physletb.2016.11.017}{\emph{Phys. Lett.}
  {\bfseries B764} (2017) 100--108},
  [\href{https://arxiv.org/abs/1512.08855}{{\ttfamily 1512.08855}}].

\bibitem{Zeng:2016sei}
X.-X. Zeng, X.-M. Liu and L.-F. Li, \emph{{Phase structure of the
  Born–Infeld–anti-de Sitter black holes probed by non-local observables}},
  \href{https://doi.org/10.1140/epjc/s10052-016-4463-4}{\emph{Eur. Phys. J.}
  {\bfseries C76} (2016) 616},
  [\href{https://arxiv.org/abs/1601.01160}{{\ttfamily 1601.01160}}].

\bibitem{Mo:2016cmi}
J.-X. Mo, G.-Q. Li, Z.-T. Lin and X.-X. Zeng, \emph{{Revisiting van der Waals
  like behavior of f(R) AdS black holes via the two point correlation
  function}},
  \href{https://doi.org/10.1016/j.nuclphysb.2017.02.015}{\emph{Nucl. Phys.}
  {\bfseries B918} (2017) 11--22},
  [\href{https://arxiv.org/abs/1604.08332}{{\ttfamily 1604.08332}}].

\bibitem{Mo:2016ijb}
J.-X. Mo, \emph{{An alternative perspective to observe the critical phenomena
  of dilaton black holes}},
  \href{https://doi.org/10.1140/epjc/s10052-017-5103-3}{\emph{Eur. Phys. J.}
  {\bfseries C77} (2017) 529},
  [\href{https://arxiv.org/abs/1607.03702}{{\ttfamily 1607.03702}}].

\bibitem{Zeng:2016aly}
S.~He, L.-F. Li and X.-X. Zeng, \emph{{Holographic Van der Waals-like phase
  transition in the Gauss–Bonnet gravity}},
  \href{https://doi.org/10.1016/j.nuclphysb.2016.12.005}{\emph{Nucl. Phys.}
  {\bfseries B915} (2017) 243--261},
  [\href{https://arxiv.org/abs/1608.04208}{{\ttfamily 1608.04208}}].

\bibitem{Zeng:2016fsb}
X.-X. Zeng and L.-F. Li, \emph{{Holographic Phase Transition Probed by Nonlocal
  Observables}}, \href{https://doi.org/10.1155/2016/6153435}{\emph{Adv. High
  Energy Phys.} {\bfseries 2016} (2016) 6153435},
  [\href{https://arxiv.org/abs/1609.06535}{{\ttfamily 1609.06535}}].

\bibitem{ElMoumni:2016eqh}
H.~El~Moumni, \emph{{Phase Transition of AdS Black Holes with Non Linear Source
  in the Holographic Framework}},
  \href{https://doi.org/10.1007/s10773-016-3197-2}{\emph{Int. J. Theor. Phys.}
  {\bfseries 56} (2017) 554--565}.

\bibitem{Li:2017gyc}
H.-L. Li, S.-Z. Yang and X.-T. Zu, \emph{{Holographic research on phase
  transitions for a five dimensional AdS black hole with conformally coupled
  scalar hair}},
  \href{https://doi.org/10.1016/j.physletb.2016.11.043}{\emph{Phys. Lett.}
  {\bfseries B764} (2017) 310--317}.

\bibitem{Li:2017xiv}
H.-L. Li and Z.-W. Feng, \emph{{Holographic Van der Waals phase transition of
  the higher dimensional electrically charged hairy black hole}},
  \href{https://arxiv.org/abs/1706.05530}{{\ttfamily 1706.05530}}.

\bibitem{Zeng:2017zlm}
X.-X. Zeng and Y.-W. Han, \emph{{Holographic Van der Waals phase transition for
  a hairy black hole}}, \href{https://doi.org/10.1155/2017/2356174}{\emph{Adv.
  High Energy Phys.} {\bfseries 2017} (2017) 2356174},
  [\href{https://arxiv.org/abs/1706.02024}{{\ttfamily 1706.02024}}].

\bibitem{Chamblin:1999tk}
A.~Chamblin, R.~Emparan, C.~V. Johnson and R.~C. Myers, \emph{{Charged AdS
  black holes and catastrophic holography}},
  \href{https://doi.org/10.1103/PhysRevD.60.064018}{\emph{Phys. Rev.}
  {\bfseries D60} (1999) 064018},
  [\href{https://arxiv.org/abs/hep-th/9902170}{{\ttfamily hep-th/9902170}}].

\bibitem{Chamblin:1999hg}
A.~Chamblin, R.~Emparan, C.~V. Johnson and R.~C. Myers, \emph{{Holography,
  thermodynamics and fluctuations of charged AdS black holes}},
  \href{https://doi.org/10.1103/PhysRevD.60.104026}{\emph{Phys. Rev.}
  {\bfseries D60} (1999) 104026},
  [\href{https://arxiv.org/abs/hep-th/9904197}{{\ttfamily hep-th/9904197}}].

\bibitem{Kubiznak:2012wp}
D.~Kubiznak and R.~B. Mann, \emph{{P-V criticality of charged AdS black
  holes}}, \href{https://doi.org/10.1007/JHEP07(2012)033}{\emph{JHEP}
  {\bfseries 07} (2012) 033},
  [\href{https://arxiv.org/abs/1205.0559}{{\ttfamily 1205.0559}}].

\bibitem{Balasubramanian:1999zv}
V.~Balasubramanian and S.~F. Ross, \emph{{Holographic particle detection}},
  \href{https://doi.org/10.1103/PhysRevD.61.044007}{\emph{Phys. Rev.}
  {\bfseries D61} (2000) 044007},
  [\href{https://arxiv.org/abs/hep-th/9906226}{{\ttfamily hep-th/9906226}}].

\bibitem{Johnson:2013dka}
C.~V. Johnson, \emph{{Large N Phase Transitions, Finite Volume, and
  Entanglement Entropy}},
  \href{https://doi.org/10.1007/JHEP03(2014)047}{\emph{JHEP} {\bfseries 03}
  (2014) 047}, [\href{https://arxiv.org/abs/1306.4955}{{\ttfamily 1306.4955}}].

\bibitem{Blanco:2013joa}
D.~D. Blanco, H.~Casini, L.-Y. Hung and R.~C. Myers, \emph{{Relative Entropy
  and Holography}}, \href{https://doi.org/10.1007/JHEP08(2013)060}{\emph{JHEP}
  {\bfseries 08} (2013) 060},
  [\href{https://arxiv.org/abs/1305.3182}{{\ttfamily 1305.3182}}].

\bibitem{Wong:2013gua}
G.~Wong, I.~Klich, L.~A. Pando~Zayas and D.~Vaman, \emph{{Entanglement
  Temperature and Entanglement Entropy of Excited States}},
  \href{https://doi.org/10.1007/JHEP12(2013)020}{\emph{JHEP} {\bfseries 12}
  (2013) 020}, [\href{https://arxiv.org/abs/1305.3291}{{\ttfamily 1305.3291}}].

\bibitem{Ryu:2006bv}
S.~Ryu and T.~Takayanagi, \emph{{Holographic derivation of entanglement entropy
  from AdS/CFT}},
  \href{https://doi.org/10.1103/PhysRevLett.96.181602}{\emph{Phys. Rev. Lett.}
  {\bfseries 96} (2006) 181602},
  [\href{https://arxiv.org/abs/hep-th/0603001}{{\ttfamily hep-th/0603001}}].

\bibitem{Bhattacharya:2012mi}
J.~Bhattacharya, M.~Nozaki, T.~Takayanagi and T.~Ugajin, \emph{{Thermodynamical
  Property of Entanglement Entropy for Excited States}},
  \href{https://doi.org/10.1103/PhysRevLett.110.091602}{\emph{Phys. Rev. Lett.}
  {\bfseries 110} (2013) 091602},
  [\href{https://arxiv.org/abs/1212.1164}{{\ttfamily 1212.1164}}].

\bibitem{PhysRevD.47.1407}
J.~D. Brown and J.~W. York, \emph{Quasilocal energy and conserved charges
  derived from the gravitational action},
  \href{https://doi.org/10.1103/PhysRevD.47.1407}{\emph{Phys. Rev. D}
  {\bfseries 47} (Feb, 1993) 1407--1419}.

\bibitem{Balasubramanian:1999re}
V.~Balasubramanian and P.~Kraus, \emph{{A Stress tensor for Anti-de Sitter
  gravity}}, \href{https://doi.org/10.1007/s002200050764}{\emph{Commun. Math.
  Phys.} {\bfseries 208} (1999) 413--428},
  [\href{https://arxiv.org/abs/hep-th/9902121}{{\ttfamily hep-th/9902121}}].

\bibitem{Mo:2016sel}
J.-X. Mo and G.-Q. Li, \emph{{Coexistence curves and molecule number densities
  of AdS black holes in the reduced parameter space}},
  \href{https://doi.org/10.1103/PhysRevD.92.024055}{\emph{Phys. Rev.}
  {\bfseries D92} (2015) 024055},
  [\href{https://arxiv.org/abs/1604.07931}{{\ttfamily 1604.07931}}].

\end{thebibliography}\endgroup
\bibliographystyle{JHEP}
\end{document}